\documentclass[twocolumn]{aastex631}
\newcommand{\s}{\ensuremath{\mbox{~s}}}

\newcommand{\cm}{\ensuremath{\mbox{~cm}}}
\newcommand{\pcmsq}{\ensuremath{\cm^{-2}}}

\newcommand{\Hii}{H\textsc{ii}}

\newcommand{\thco}{$^{13}$CO}

\newcommand{\etco}{C$^{18}$O}
\newcommand{\vel}{km\,s$^{-1}$}

\newcommand{\imcoorco}{$\alpha_{2000}=18^{\mathrm{h}}53^{\mathrm{m}}17\fs989,\ \delta_{2000}=+1\degr25\arcmin24\farcs18$}
\newcommand{\imcoorhco}{$\alpha_{2000}=18^{\mathrm{h}}53^{\mathrm{m}}18\fs585,\ \delta_{2000}=+1\degr26\arcmin13\farcs12$}

\newcommand{\msun}{$M_{\odot}$}
\newcommand{\lsun}{$L_{\odot}$}
\newcommand{\um}{$\mu$m}
\newcommand{\cmcm}{cm$^{-2}$}
\newcommand{\egcite}{\citep[e.g.,][]}

\newcommand{\hii}{H\textsc{ii}}

\newcommand{\filname}{G34}
\newcommand{\mdotyr}{$M_{\odot}$~yr$^{-1}$}

\newcommand{\hco}{HCO$^+$}
\newcommand{\hthco}{H$^{13}$CO$^+$}
\newcommand{\hthcn}{H$^{13}$CN}

\usepackage{ae,aecompl}
\usepackage{subfigure}
\usepackage{graphicx}	% Including figure files
\usepackage{amsmath}	% Advanced maths commands
\usepackage{amssymb}	% Extra maths symbols
\usepackage{color,xcolor}
\usepackage{float}
\def   \apj {{\rm {ApJ}}}

\def   \apjs {{\rm {ApJS}}}
\def   \apjl {{\rm {ApJL}}}

\def   \aap {{\rm {A\&A}}}

\def   \mnras {{\rm {MNRAS}}}

\graphicspath{{0figs/}}

\received{xxx}
\revised{xxx}
\accepted{xxx}
\shorttitle{Multiscale scenario of high-mass star formation}
\shortauthors{Liu et al.}

\begin{document}
\title{\bf Multiscale dynamical scenario of high-mass star formation in an IRDC filament G34}

\submitjournal{ApJ}
\correspondingauthor{Hong-Li Liu, Sirong Pan, Sheng-Li Qin}
\email{hongliliu2012@gmail.com, siyun.pan@foxmail.com, qin@ynu.edu.cn}

\author{Sirong Pan*}
\affiliation{School of physics and astronomy, Yunnan University, Kunming, 650091, PR China}

\author{Hong-Li Liu*}
\affiliation{School of physics and astronomy, Yunnan University, Kunming, 650091, PR China}

\author{Sheng-Li Qin*}
\affiliation{School of physics and astronomy, Yunnan University, Kunming, 650091, PR China}

\begin{abstract}
There is growing evidence that high-mass star formation (HMSF) is a multiscale, dynamical process in molecular clouds, where filaments transport gas material between larger and smaller scales. We analyze here multiscale gas dynamics in an HMSF filamentary cloud, G034.43+00.24 (G34), using APEX observations of \etco~(2-1), \hco/\hthco~(3-2), and HCN/\hthcn~(3-2) lines. We find large-scale, filament-aligned velocity gradients from \etco\ emission, which drive filamentary gas inflows onto dense clumps in the middle ridge of G34. The nature of these inflows is  gravity-driven.
We also find clump-scale gas infall in the middle ridge of MM2, MM4, and MM5 clumps from other lines. Their gas infall rates could depend on large-scale filamentary gas inflows since the infall/inflow rates on these two scales are comparable. 
We confirm that the multiscale, dynamical HMSF scenario is at work in G34. 
 It could be driven by gravity up to the filament scale, beyond which turbulence originating from several sources including gravity could be in effect in G34.
\end{abstract}


\section{Introduction}
High-mass  star formation (HMSF, for $M_{*}>8$\,\msun) remains elusive in astrophysics. Two major theories, `core accretion' \citep{McK03} and `competitive accretion' models \citep{Bon01} have competed with each other for nearly two decades. The `core accretion' model suggests that high-mass stars form from  the monolithic collapse of massive cores, in a similar way to low-mass stars. The `competitive accretion' model contends that high-mass stars form from the initial fragmentation of massive clumps into a cluster of smaller fragments (e.g., cores), which then compete for gas from a common reservoir. Both types of models did not pay much attention on the important role of large-scale interstellar medium (ISM) environments beyond clump/core scales, which can influence the gas infall onto young stars, and thus determine their final mass.  This situation could be due to the computational limitation at that time. 

This critical role has been gradually revealed in recent multiscale, high-resolution observations \egcite{Per13, Yua18, Avi21,  Liu23, Yan23, Xu 23, HeY23} and large-scale dynamical theoretical models, such as ``global hierarchical collapse'' (GHC, \citealt{Vaz19, Vaz23}) and ``inertial-inflow'' (I2, \citealt{Pad20}). For example, \citet{Yan23} observed for the first time direct evidence of a multiscale, dynamical gas infall process in the G310 cloud, which is an elegant massive hub-filament system, and thus argued that HMSF therein can be described by a multiscale mass accretion/transfer scenario, from large-scale filamentary clouds through clumps down to smaller-scale cores. Also, such observational evidence provides strong support to the latest generation of GHC and I2 models. They advocate for a multiscale HMSF process  that involves fragmentation and gas infall onto protostars over all hierarchical density scales of molecular clouds. Moreover, both models agree on the gravity-driven gas-infall/accretion on small scales (e.g. cores). However, they differ on the origin of large-scale gas inflows, where the GHC model favors a gravity-driven hierarchical gas infall while the I2 model supports a turbulence-driven gas inflow. This physical difference between theoretical models strongly needs to be addressed through dedicated multiscale kinematic and/or dynamical studies toward HMSF regions.

\begin{figure}
\centering
\includegraphics[width=0.4\textwidth, height=0.8\textwidth]{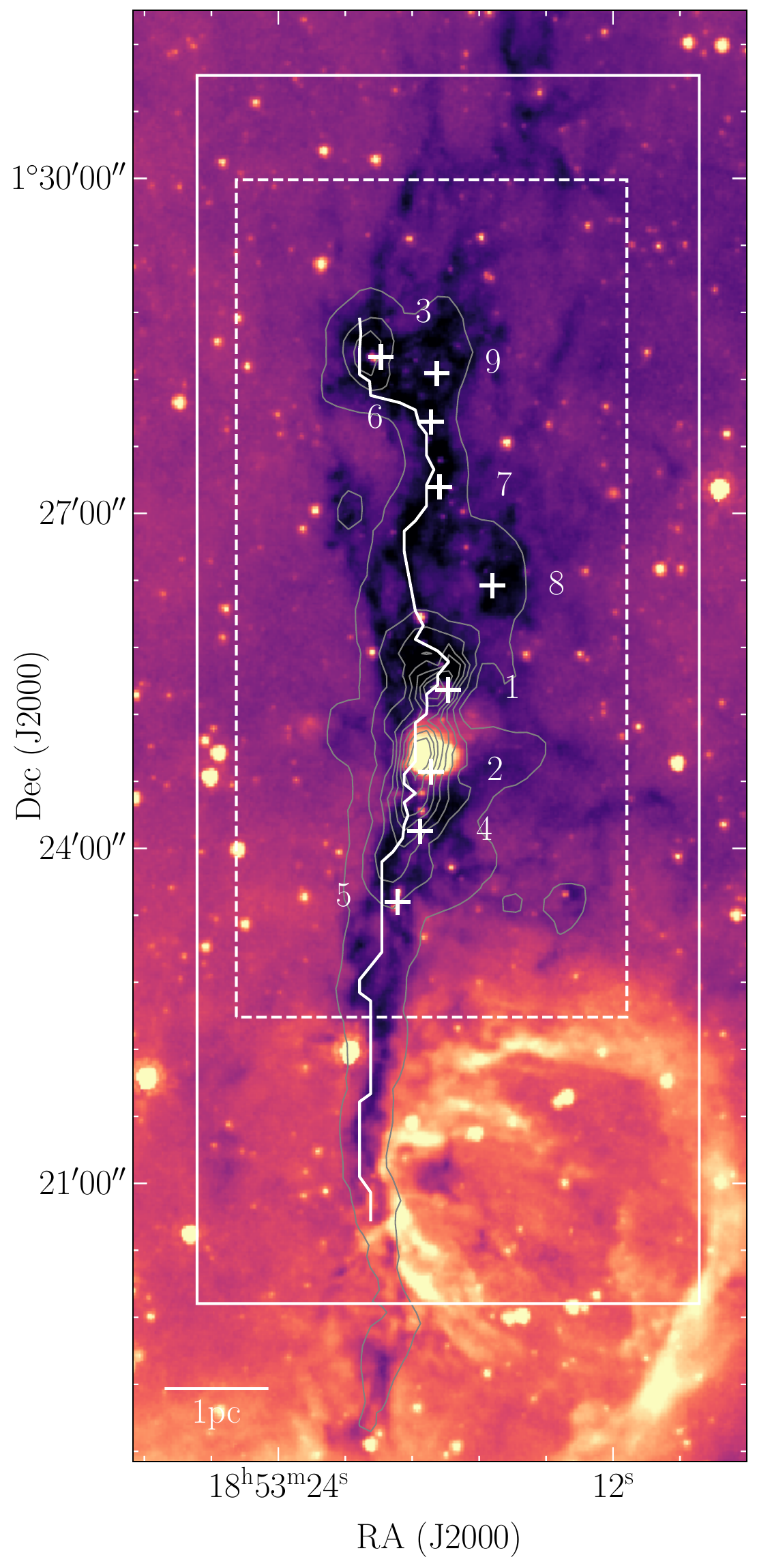}
\caption{{\it Spitzer} 8\,\um\ image (color-scale) of the G34 IR-dark cloud overlaid
with contours of the column density ($N$(H$_2$)). The contours start at 2\,$N_{\rm bg}$ increasing in steps following the power-law $D=3\times L^{p}+2$, where the dynamical range, $D$, is defined as the ratio of the $N$(H$_2$) peak  to the background value,  $L$ is the level number ($L=[1,8]$ in this case), and the index $p$ is derived from both $D$ and the maximum $L$. The curve delineates the filament's ridgeline. while plus symbols indicate nine dense clumps of \citet{Rat06}. The solid- and dashed- line boxes represent two mapping areas of APEX observations (see Sect.\,2).}
\label{fig:filoverview}

\end{figure}

We present here multiscale kinematic/dynamical analysis toward a most-studied, filamentary HMSF cloud, G034.43$+$00.24 (hereafter \filname). The major goal is to investigate the physical origin of the large-scale gas infall/inflow motions, which is a key debate in the latest generation of HMSF models. Located at $\sim 3.7$\,kpc\footnote{The G34 cloud has an updated distance of $\sim 3$\,kpc from new measurements of the trigonometric parallax of 22\,GHz water masers \citep{Mai23}. To maintain the consistency with our previous work \citep{Liu20, Liu22a, Liu22b}, we still use here the commonly-adopted distance of $\sim 3.7$\,kpc.},
the G34 cloud is still in early evolutionary stages of HMSF, as reflected from the infrared (IR)-dark appearance in {\it Spitzer}-8\,\um\ emission (see Fig.\,\ref{fig:filoverview}) across the cloud. Moreover, the cloud hosts nine dense clumps (i.e., MM1 to MM9, \citealt{Rat06}), 
% identified from $11\arcsec$-resolution 1.2\,mm observations by \citet{Rat06}. 
four of which are HMSF sites as characterized by an associated ultra-compact \Hii\ region in MM2, and high bolometric luminosities ($9,000-32,000$\,\lsun) in MM1, MM3 and MM4 \citep{Rat05}.  In this paper, we report the finding of the filament- and clump-scale mass flows, and place an observational constraint on the origin of the multiscale gas flows driving HMSF in G34.

\begin{figure*}
\centering
\includegraphics[width=0.46\textwidth, height=0.98\textwidth]{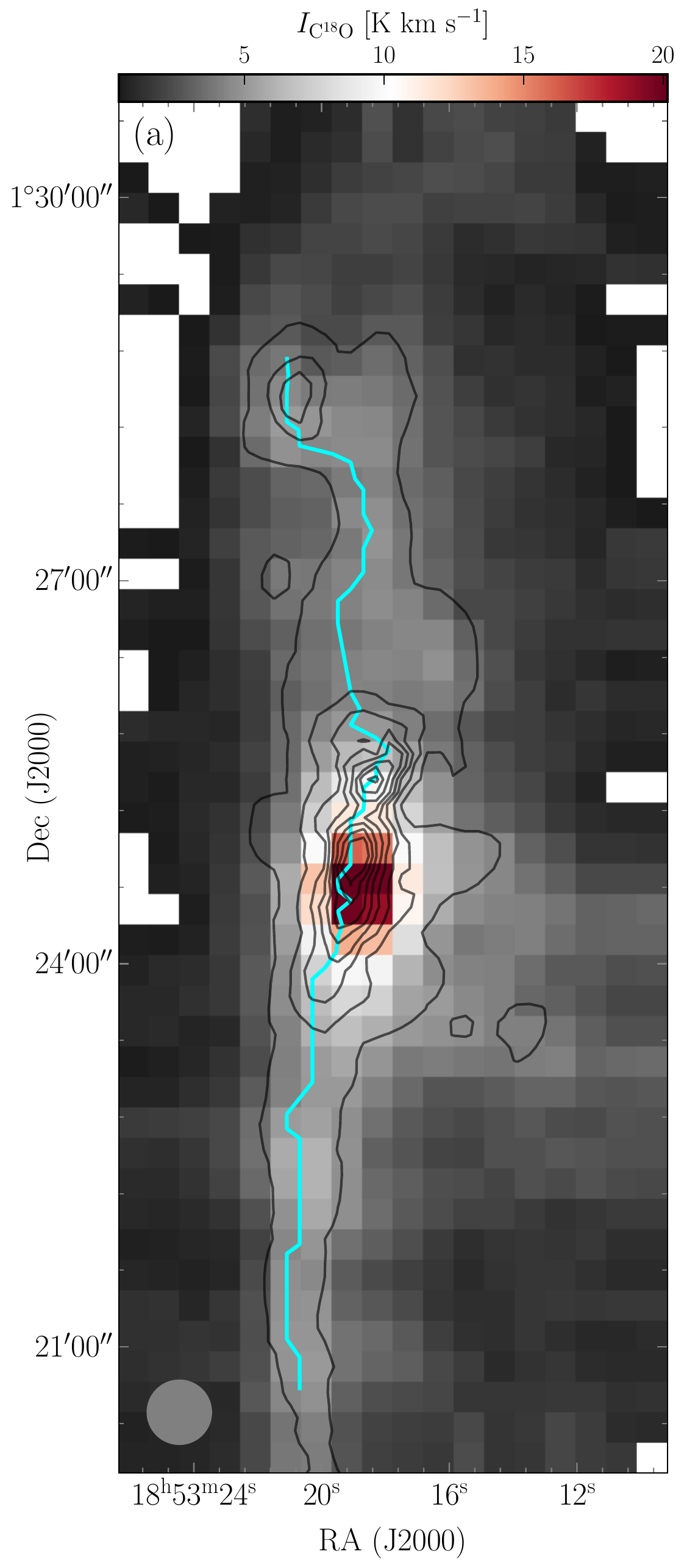}
\includegraphics[width=0.46\textwidth, height=0.98\textwidth]{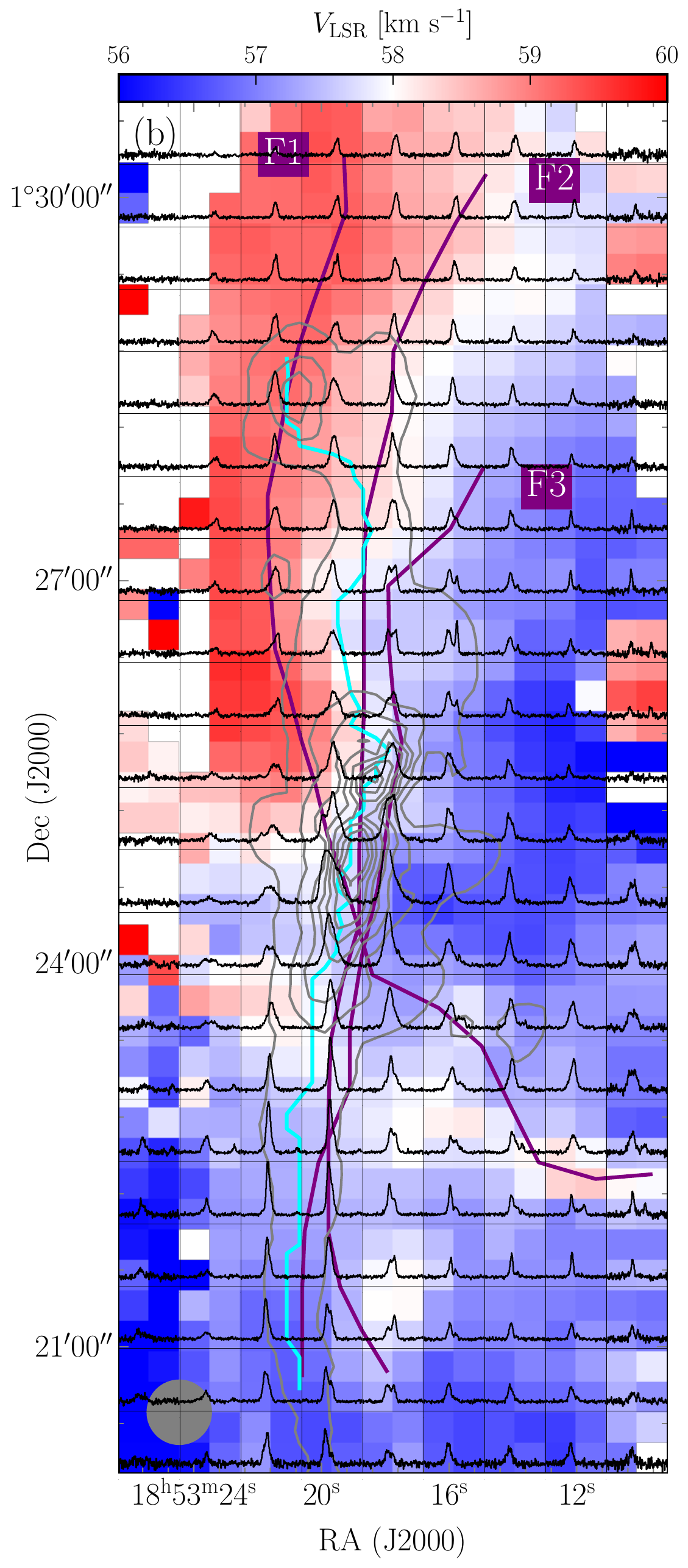}
\caption{ Maps of Moment\,0 (panel\,a) and Moment\,1 (panel\,b) overlaid with the contours of the column density ($N$(H$_2$)). They are the same as in Fig.\,\ref{fig:filoverview}. The cyan curve in both panels indicates the ridgeline of the major filament derived from the $N$(H$_2$) map while the purple ones only in panel\,b outline the three subfilaments, F1--F3, visually identified from the \etco~(2--1) cube.
}
\label{fig:m1_c18O}
\end{figure*}

\section{Observations and data reduction}

We observed  several major molecular lines between 214.0 and 272.5\,GHz  toward G34  in October 2019 using the Atacama Pathfinder Experiment (APEX) telescope \citep{Gus06}. The observations include three observing setups, {\it Setups-A, B}, and C. The observing frequencies were tuned to be
at 220\,GHz for {\it Setup-A}  primarily cover CO, \thco, and \etco~(2-1) at an angular resolution of $\theta \sim 28\arcsec$,  at 267\,GHz for {\it Setup-B} to cover \hco\ and HCN~(3-2) at $\theta \sim 24\arcsec$, and at 255\,GHz {\it Setup-C} to cover \hthco\ and \hthcn~(3-2) at $\theta \sim 23\arcsec$. {\it Setups-A/B} were configured in
an on-the-fly mode to map a large spatial extent of the G34 filament, which is $3\farcm5 \times 10\farcm0$ for {\it Setup-A} centered at \imcoorco, and $2\farcm5 \times 6\farcm5$ for {\it Setup-B} centered at \imcoorhco\ (see  solid, and dashed-line boxes in Fig.\,\ref{fig:filoverview}), while {\it Setup-C} was designed in a position-switching mode to target nine dense clumps.  We processed the raw data to cubes using the IRAM’s GILDAS software \citep{Gui00}. The final data cubes have a typical rms of $\sim 0.1$\,K for {\it Setups-A/B},  and  0.03\,K for {\it Setup-C} at a velocity resolution of 0.1\,\vel. The details about the observations and data reduction can be found in \citet{Liu20}. 

In this paper, we primarily focus on the lines of \etco~(2-1) from {\it Setup-A}, \hco/\hthco~(3-2), and HCN/\hthcn~(3-2) from {\it Setups-B/C}. For the \etco~(2-1) line, we followed the approach adopted by \citet{Pan23} and \citet{Yan23} to produce a de-noised/modelled data cube using the BTS algorithm \citep{Cla18}, which reduces the noise effect of the observed data on subsequent line emission analysis.  Based on the average spectrum of \etco\ over the entire G34 cloud, the velocity range of the modelled cube was set in [55, 62]\,\vel. In this range, the fraction of one and two velocity components in the total modelled spectra is 83\%, and 14\%, respectively. This indicates that \etco~(2-1) line emission of the G34 cloud is characterized mostly as a single velocity component, and thus rather simple for kinematic analysis. Likewise, the de-noised data cube was made for the \thco~(2-1) line, showing very complicated line emission with more than two velocity components over 96\% in area of the observed cloud. This data cube will therefore not be considered hereafter.

\section{Results and analysis}
\begin{figure}
\centering
\includegraphics[width=0.45\textwidth, height=0.4\textwidth]{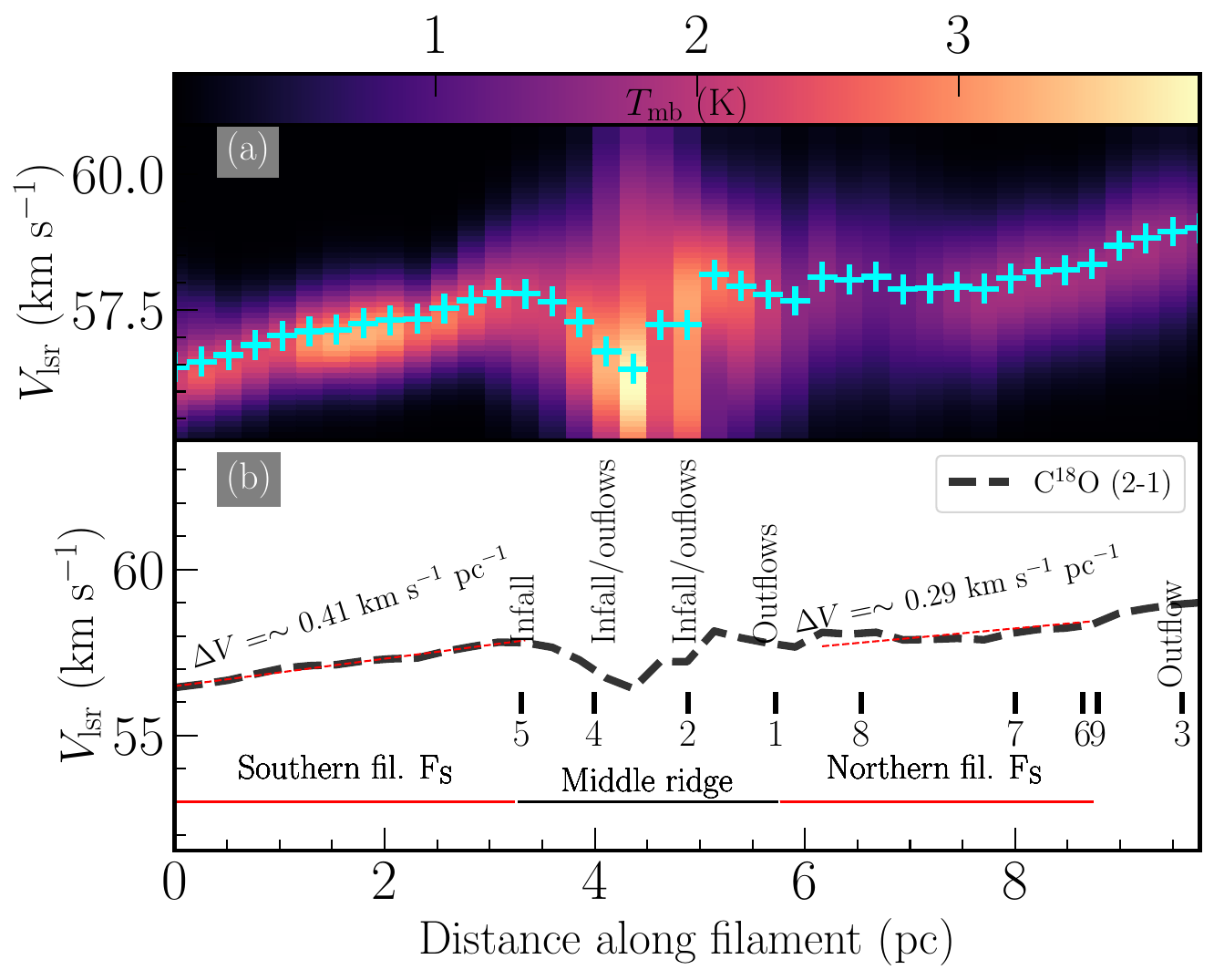}
\includegraphics[width=0.427\textwidth, height=0.20\textwidth]{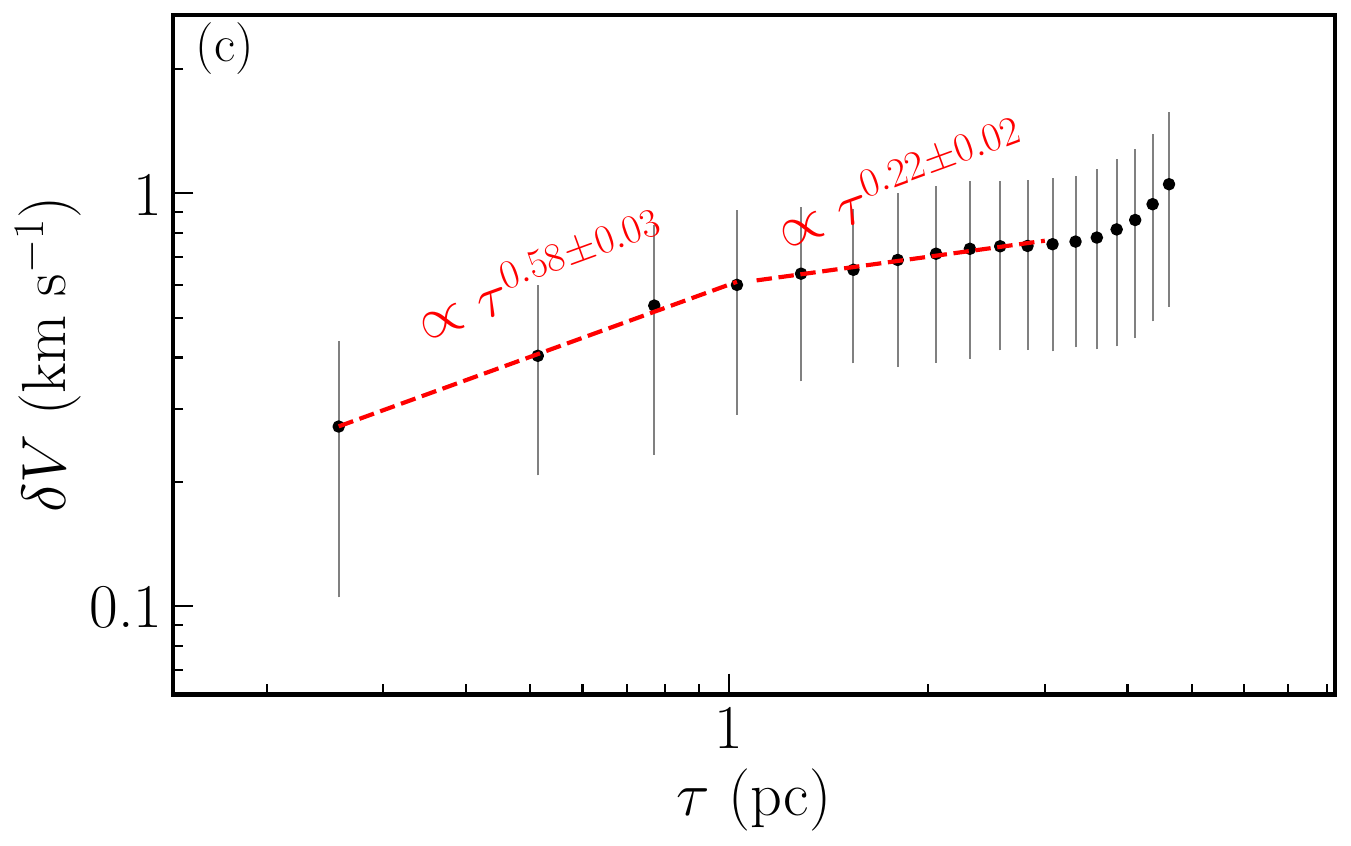}
\caption{(a): PV diagram of \etco~(2-1) extracted within [50, 65]\,\vel along the filament enclosed within 0.6\,pc, starting from the southern end of the ridgeline to its northern end. The average velocities weighted by line brightness temperature are indicated in crosses, which represent the longitudinal velocity distribution along filament. (b): fitting to the longtitudinal velocity distribution. Nine dense clumps are labeled, some of which are associated with either infall or outflows (see text).
(c):  second-order structure function analysis for the longitudinal velocity distribution. The segmented power-law fits are applied to describe the scale-dependent velocity fluctuation (i.e., $\delta V \propto \tau^{\gamma}$) on the scale $\tau$ regimes below and above 1\,pc.  The error bar corresponds to the standard error of the mean.
}
\label{fig:pv_diagram}
\end{figure} 

\subsection{Cloud-scale gas emission}
 To overview large-scale, cloud gas emission, we make use of \etco~(2--1) emission. Figure\,\ref{fig:m1_c18O}a presents the moment\,0 map of the \etco\ line. For comparison, the column density ($N{\rm (H_2)}$) map, whose calculation can be found below, is shown in contours. \etco~(2--1) emission is detectable with a level of $\geq3$\,rms (1\,rms$=0.1$\,K) in 68\% of the whole observed area, and matches spatially the IR-dark appearance at 8\,\um\ very well (see Fig.\,\ref{fig:filoverview}). As shown in Fig.\,\ref{fig:m1_c18O}, the same spatial match can be found with the $N{\rm (H_2)}$ distribution, where   strong line emission peaks representative of bright gas emission correspond to the distribution of high column densities along the G34 filament. 

Fig.\,\ref{fig:m1_c18O}b displays the map grid of spectra of \etco\ overlaid on its moment\,1 map created over a velocity range of [55, 62]\,\vel. 
The moment\,1 map presents the centroid (mean) velocity distribution of \etco\ of the entire G34 cloud  that hosts the major dense filament. As mentioned earlier, a small fraction of positions have spectra composed of at least two velocity components, which can also be reflected at several positions from the grid map of \etco\ spectra in Fig.\,\ref{fig:m1_c18O}. To minimize the effect of the multiple velocity components on the determination of the centroid velocity, we calculated the moment\,1 map using the intensity-weighted approach by each component according to \citet{Pan23}.  As a result, the global velocity distribution in Fig.\,\ref{fig:m1_c18O} shows an east-west velocity gradient across the G34 filament in its north and a roughly flat distribution in the south, which agrees with the finding from N$_2$H$^{+}$~(1--0) emission by \citet{Tan19}. 

 To explore the potential nature of the east-west velocity gradient across the northern filament, we created a velocity channel map with a channel width of 1.0\,\vel, as depicted in Fig.\,\ref{fig:vel_chan} of Appendix. This map reveals at least three velocity-coherent subfilaments, labeled as F1--F3, which were identified by eye. Their identification requires a refinement when conducting detailed analysis such as gas dynamics within subfilaments, which is beyond the scope of this paper. Their elongations align with the dust lanes visible in 2\arcsec-resolution 8\,\um\ emission. F3 appears to be separated from F1 and F2 by more than two beams of \etco~(2--1) observations, while F1 and F2 cannot be spatially resolved within a beam. It's worth noting that these subfilaments cannot be identified from the velocity-integrated intensity map of \etco, possibly due to the selection effect of the current line being more sensitive to envelope diffuse gas than to the interior dense gas of the cloud. The systemic velocity of F1--F3 is centered at approximately $\sim 57.0$, $58.0$, and 59.5\,\vel, respectively, indicating an increasing velocity trend from F1 to F3. This trend aligns with the east-west velocity gradient observed from the moment\,1 map across the northern G34 cloud (see Fig.\,\ref{fig:m1_c18O}). However, the relatively flat velocity distribution over the southern cloud suggests that the southern parts of both F1 and F2 subfilaments could share a consistent velocity trend.

\subsection{Filament-scale longitudinal gas dynamics}
\label{sec:dynamics_cloud} 
We constructed from the \etco\ de-noised cube the position-velocity (PV) diagram along the filament ridgeline  within a $0.6$\,pc scale that represents the filament's width (see below).  The ridgeline was extracted from the $18\arcsec$-resolution H$_2$ column density ($N{\rm (H_2)}$) map by identifying the peak value along the filament (see Fig.\,\ref{fig:filoverview}, and Appendix\,\ref{app:radial_profile}).  Following \citet{Pan23,Yan23}, the $N{\rm (H_2)}$ map was created from {\it Herschel} data. For simplicity, the ratio of 160, 250\,\um\ images was used as a dust temperature ($T_{\rm d}$) tracer and the latter was then used together with the 250\,\um\ image to derive the $N{\rm (H_2)}$ map assuming an index of $\beta=1.75$ for dust opacity \egcite{Per23}. 

Figure\,\ref{fig:pv_diagram}a shows the PV diagram along the filament overlaid with the intensity-weighted average velocity (in plus symbol). The diagram was made from  the southern end of the ridgeline to its northern end.
Overall, the PV diagram shows two coherent velocity gradients, one from the southern end up to the MM5 clump that corresponds to the $\sim3.25$\,pc-long southern filament $F_{\rm S}$ (see Fig.\,\ref{fig:filoverview} and Fig.\,\ref{fig:pv_diagram}b), and the other one in the north between MM1 to MM9 that corresponds to the $\sim 3$\,pc-long northern filament $F_{\rm N}$. In addition, a wave-like pattern of velocity distribution can be found in the middle ridge of the filament that hosts the MM1, MM2, MM4 and MM5 clumps. Moreover, in the  $F_{\rm S}$ and  $F_{\rm N}$ filaments the velocities are distributed within $5$\,\vel\ while in the middle ridge they spread over $5$\,\vel\ up to more than $17.5$\,\vel. The latter widespread distribution could result either from  outflow feedback (i.e., MM1, MM2, MM4 in Fig.\,\ref{fig:pv_diagram}b, \citealt{San10,Rat05,Liu20}), or from converging flows by several subfilaments around the middle ridge as reflected from filamentary branches at 8\,\um\ emission. We performed two linear fits to the velocity gradient ($\nabla V$) for the $F_{\rm S}$ and  $F_{\rm N}$  filaments 
(see Fig.\,\ref{fig:pv_diagram}b), yielding
$\nabla V_{\rm fs} = \sim0.4$\,\vel\,pc$^{-1}$,  and $\nabla V_{\rm fn} = \sim0.3$\,\vel\,pc$^{-1}$, respectively.  It is worth noting that subfilaments F1--F3 are not considered individually in analysis of velocity gradients, and instead are treated as a whole in terms of the southern $F_{\rm S}$ and northern $F_{\rm N}$ filaments. 
The two filament-aligned velocity gradients could be related to filamentary gas inflows onto the middle ridge of clumps (i.e., MM1, MM2, MM4 and MM5) in G34. 
Note that the F1--F3 subfilaments have a good spatial association with 8\,\um\ IR-dark lanes that appear to form an HFS strucrure with the central hub around the massive MM1 clump (see Figs.\,\ref{fig:filoverview}, and \ref{fig:m1_c18O}b). In the context of the HFS morphology, the hub-composing filamentary gas (e.g., F1--F3) can be expected to inflow onto the central massive hub clump.
In addition, the $F_{\rm N}$-aligned velocity gradient may correspond to the  increasing velocity trend among subfilaments from F3 to F1 that could be a result of the merging of the subfilaments. This contribution may not be significant since the $F_{\rm N}$ ridge line does not follow completely the maximum of that increasing velocity trend, as shown in Fig.\,\ref{fig:m1_c18O}.

 If southern and northern filaments are guiding gas inflows onto the middle ridge of G34, we can gauge the related inflow rate from the filament-rooted velocity gradients by the relation $\dot{M_{\rm f}} = \frac{M_{\rm f} \nabla V_{\rm f}}{\rm tan(\phi)}$ for an inclination angle of the filament to line of sight, $\phi$ \citep{Kir13,Yan23}. 
The mass ($M_{\rm f}$) of the  $F_{\rm S}$ and  $F_{\rm N}$ filaments can be estimated from the $N{\rm (H_2)}$ map along the filament enclosed in the $\sim0.6$\,pc width, having a total mass of $\sim 550$\,\msun, and $\sim930$\,\msun, respectively. This estimate accounts for the subtraction of background emission of $N{\rm bg}=2.7\times10^{22}$\,\cmcm. The $\sim0.6$\,pc size is the average width of the whole G34 filament, inferred from its average radial $N{\rm (H_2)}$ profile, which is shown in Figure\,8 of Appendix, and referred to \citet{Liu18,Pan23}. Assuming $\phi \sim 45\degr$, the gas inflow rate is estimated to be $\sim 1.6\times10^{-4}$\mdotyr\ for the $F_{\rm S}$ filament, and $\sim 3.9\times10^{-4}$\mdotyr\ for $F_{\rm N}$. In turn, the total inflow rate of both filaments toward the middle ridge of G34 is  $\sim 5.5\times10^{-4}$\,\mdotyr. This value should be a lower limit, and would be doubled to approach an order of $10^{-3}$\,\mdotyr\ if the remaining gas inflowing off the two $F_{\rm S}$ and $F_{\rm N}$ filaments were considered \citep{Per13, Yan23}.

\subsection{Clump-scale infall motions}
\label{sec:dynamics_clump}
\begin{figure}
\centering
\includegraphics[width=0.232\textwidth, height=0.35\textwidth]{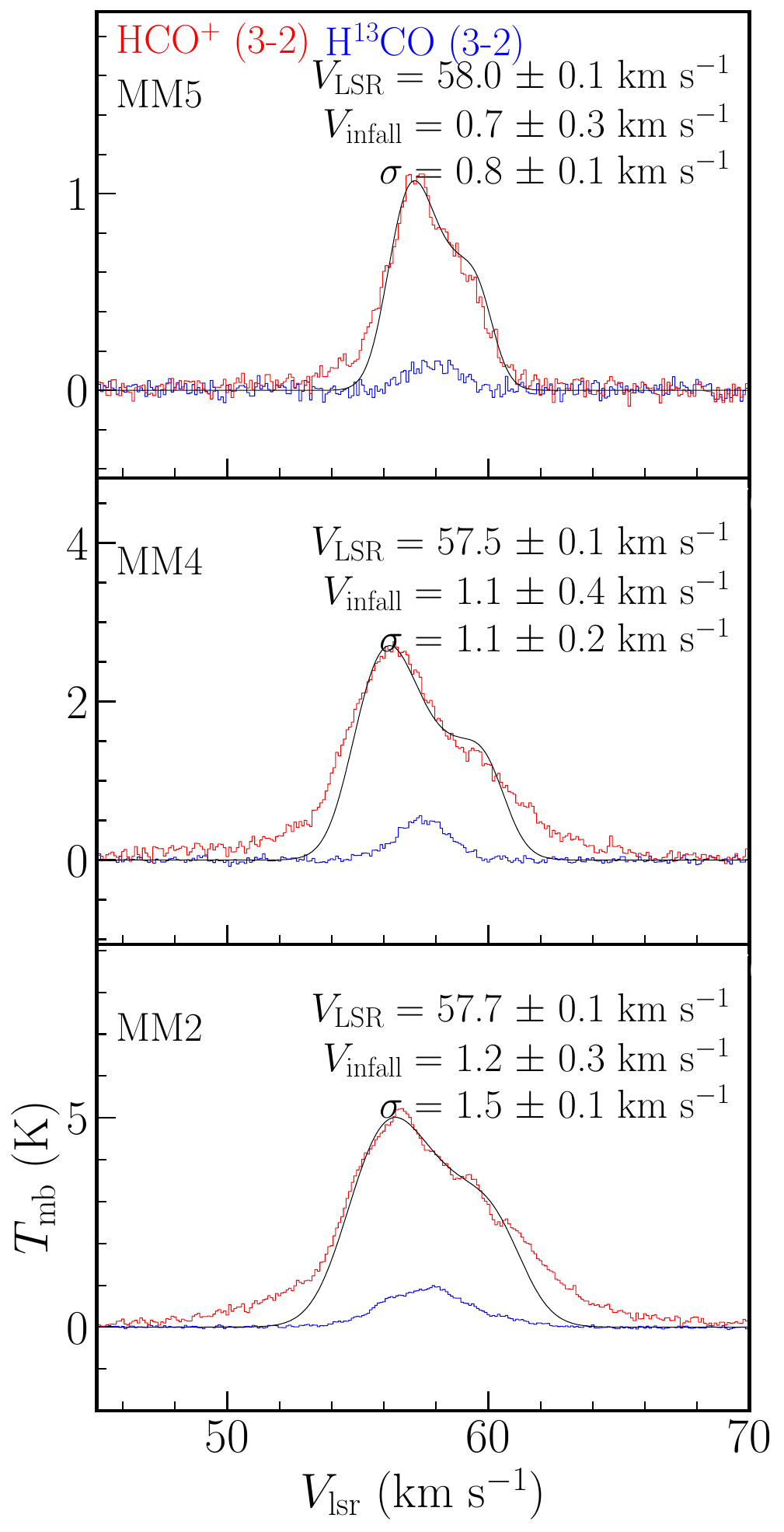}
\includegraphics[width=0.232\textwidth, height=0.35\textwidth]{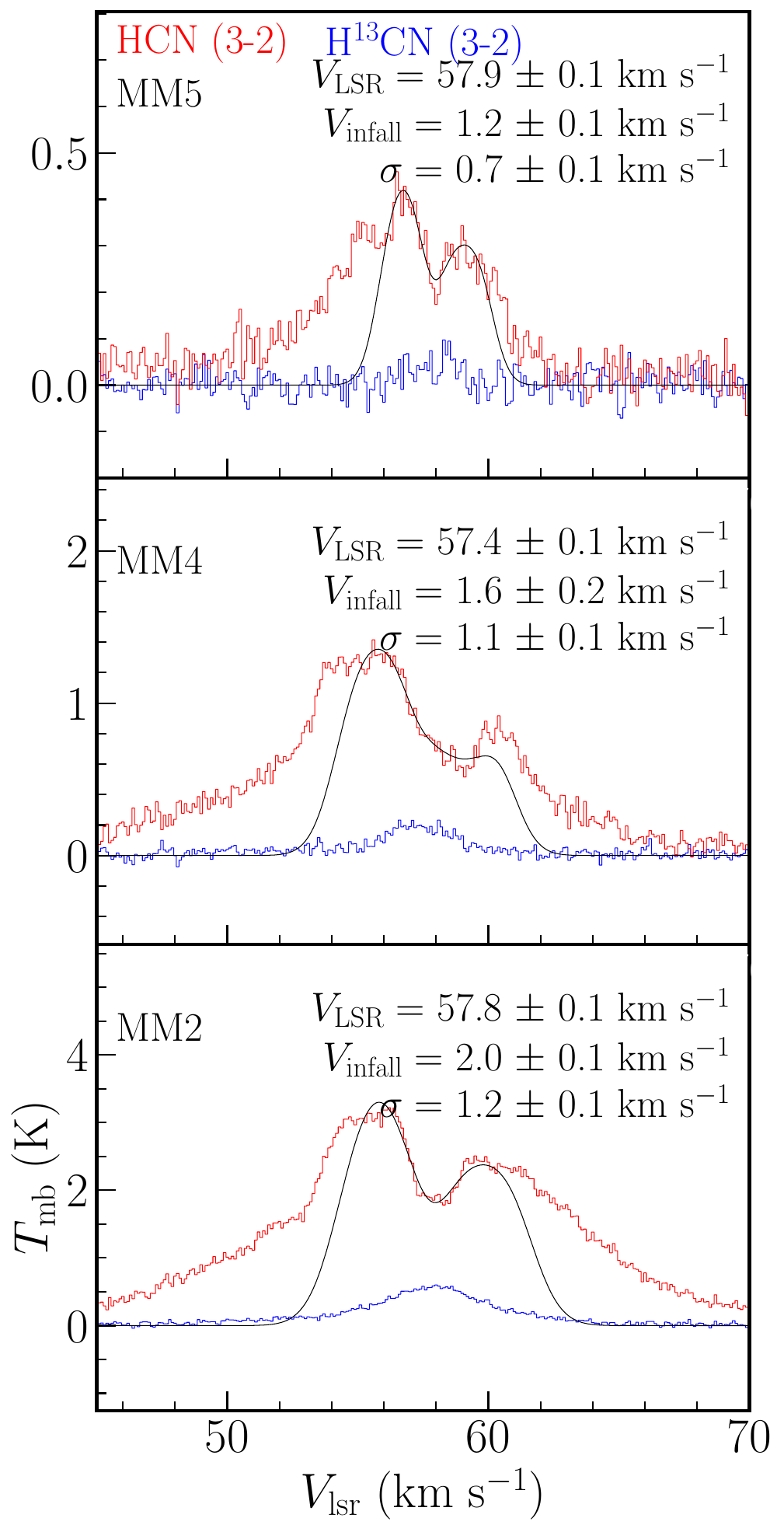}
\caption{Fitting of gas infall model toward MM2, MM4, and MM5 clumps. The red and blue curves stand for observed spectra  while the black curve for the fitting result.  The fitted parameters are labelled on top right of each panel. 
}
\label{fig:spectra_infall}
\end{figure}

Clump-scale dynamics, such as gas infall and outflow motions, are common indicators of star formation. 
We combine two optically-thick/thin pairs of the \hco/\hthco~(3--2), and HCN/\hthcn~(3--2) lines to cross-identify infall signatures associated with dense clumps in G34.  The velocity-integrated intensity maps of both \hco\ and HCN (see Fig.\,\ref{fig:M0_h13cop} for the intensity map of \hco\ as an example) indicate very strong correspondence between dense gas emission and compact dust continuum emission over the MM1--MM5 clumps, which enables the search for clump-scale gas infall motions.
A double-peaked or asymmetric \hco/HCN line profile with brighter emission on the blue side of the clump's systemic velocity is generally interpreted as gas infall if the \hthco/\hthcn\ profile is single peaked at the systemic velocity. Following this standard and inspecting the clump-averaged spectra, we find three infall signatures associated with the MM2, MM4, and MM5 clumps (see Fig.\,\ref{fig:spectra_infall}). 

We evaluated the clump-scale infall velocity by fitting the ``Hill5" model \citep{De 05} to the asymmetric line profiles of \hco\ and HCN. It is a simple radiative transfer model mimicking a collapsing system. 
Figure\,\ref{fig:spectra_infall} shows the fitting results for  the MM2, MM4, and MM5 clumps, where the black curve represents the fitted spectrum. Evidently, the model with an only infall component cannot fully reproduce the observed spectra and an additional component, such as outflows, are required to account for broad line wings, especially for MM2 and MM4. These two clumps are indeed observed to have associated  outflows \egcite{San10,Rat05,Liu20}. Regardless of the broad line wings, the central asymmetric profiles of \hco\ and HCN responsible for infall are fitted very well. The fitted parameters are indicated in each panel of Fig.\,\ref{fig:spectra_infall}, including the systemic velocity, $V_{\rm LSR}$, infall velocity, $V_{\rm inf}$, and velocity dispersion, $\sigma$. 

We calculated the virial parameter of the MM2, MM4 and MM5 clumps as $\alpha_{\rm vir}=\frac{5\sigma^2 R_{\rm clp}}{{\rm G}M_{\rm clp}}$ \egcite{Kru05,Liu19} where G is the gravitational constant. The mass ($M_{\rm clp}$) within the clump radius ($R_{\rm clp}$) was recalculated from the $N{\rm (H_2)}$  map with background emission subtracted. 
For a consistency with our previous work (\citealt{Liu20}), the clump radii determined by  \citet{Rat06} are adopted. Overall,  the clump radii of MM2 and MM4 are almost the same at 0.23\,pc while MM5's radius is about twice as large. However, the latter could not be well determined since 1.2\,mm dust emission in \citet{Rat06} around MM5 is too flat/diffuse to well define the clump. The MM5's radius is therefore assume to be the same as 0.23\,pc. As a result, the mass of MM2, MM4, and MM5 is estimated to be $\sim 777$\,\msun, 428\,\msun, and 96\,\msun, respectively. Furthermore, the velocity dispersion of the three clumps in order, as indicated in Fig.\,\ref{fig:spectra_infall} yields viral parameters of $\alpha_{\rm vir}=\sim 0.8$, 0.7, 1.7, respectively. These values satisfy $\alpha_{\rm vir}^{\rm crit}\leq2$ for gravitational collapse \egcite{Kau13}, which is consistent with the infall signatures observed in the three clumps.

Moreover, we estimated the clump-scale gas infall rate of the three clumps using the relation $\dot{M_{\rm clp}} = M_{\rm clp} V_{\rm inf} / R_{\rm clp}$.  The infall velocity ($V_{\rm inf}$),  given by the infall model (see Fig.\,\ref{fig:spectra_infall}), leads to an infall rate of  $5.7\pm1.4\times10^{-3}$\,\mdotyr\ for MM2, $2.7\pm0.6\times10^{-3}$\,\mdotyr\ for MM4, and $4.4\pm1.3\times10^{-4}$\,\mdotyr\ for MM5, where the uncertainties are propagated from those of $V_{\rm inf}$.  For MM2, our result is consistent with the value of $1.8\times10^{-3}$\,\mdotyr\ by \citet{San10}. For MM4 and MM5, their gas infall rates have been reported here for the first time.

\section{Discussion}
\subsection{Multiscale, dynamical HMSF scenario in \filname}
\begin{figure*}
\centering
\includegraphics[width=0.32\textwidth, height=0.23\textwidth]{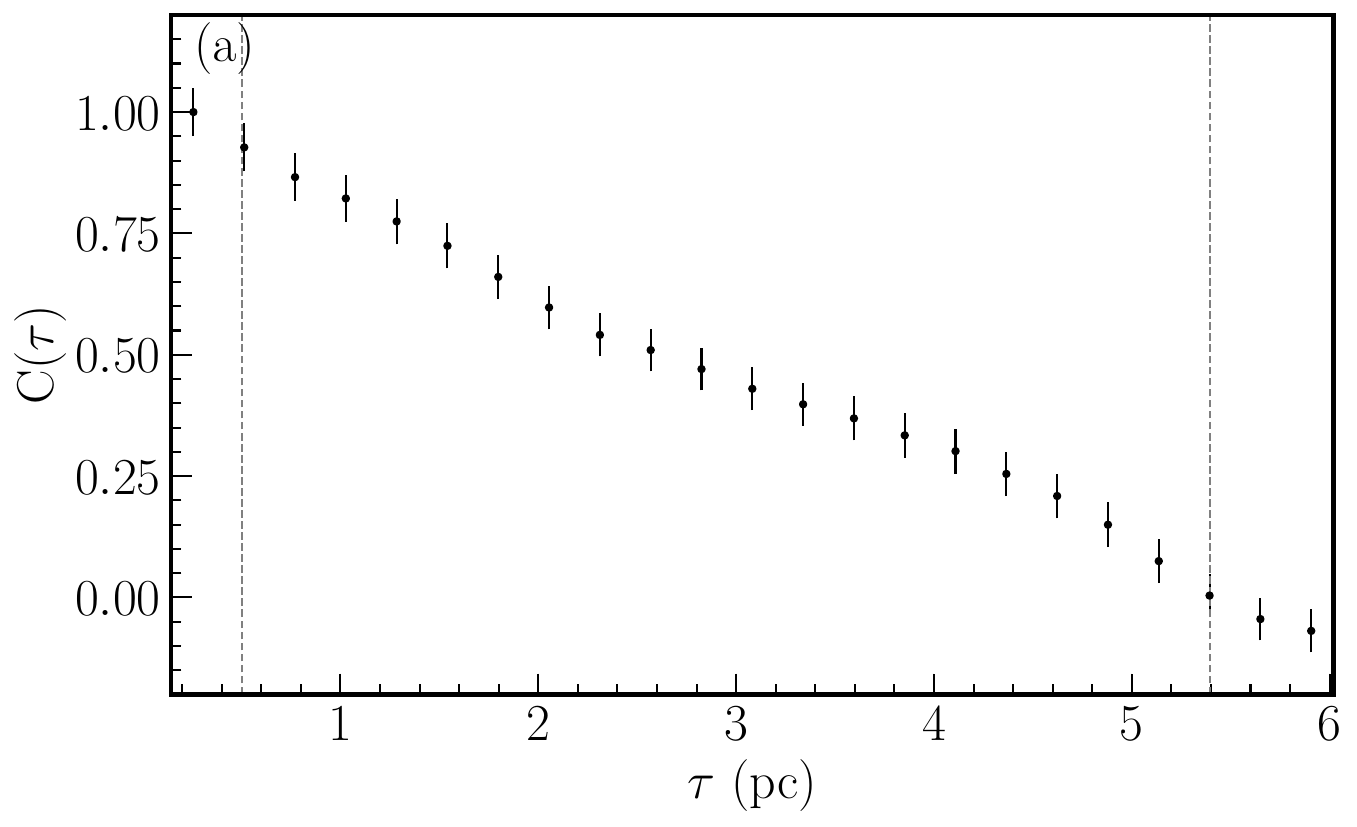}
\includegraphics[width=0.32\textwidth, height=0.23\textwidth]{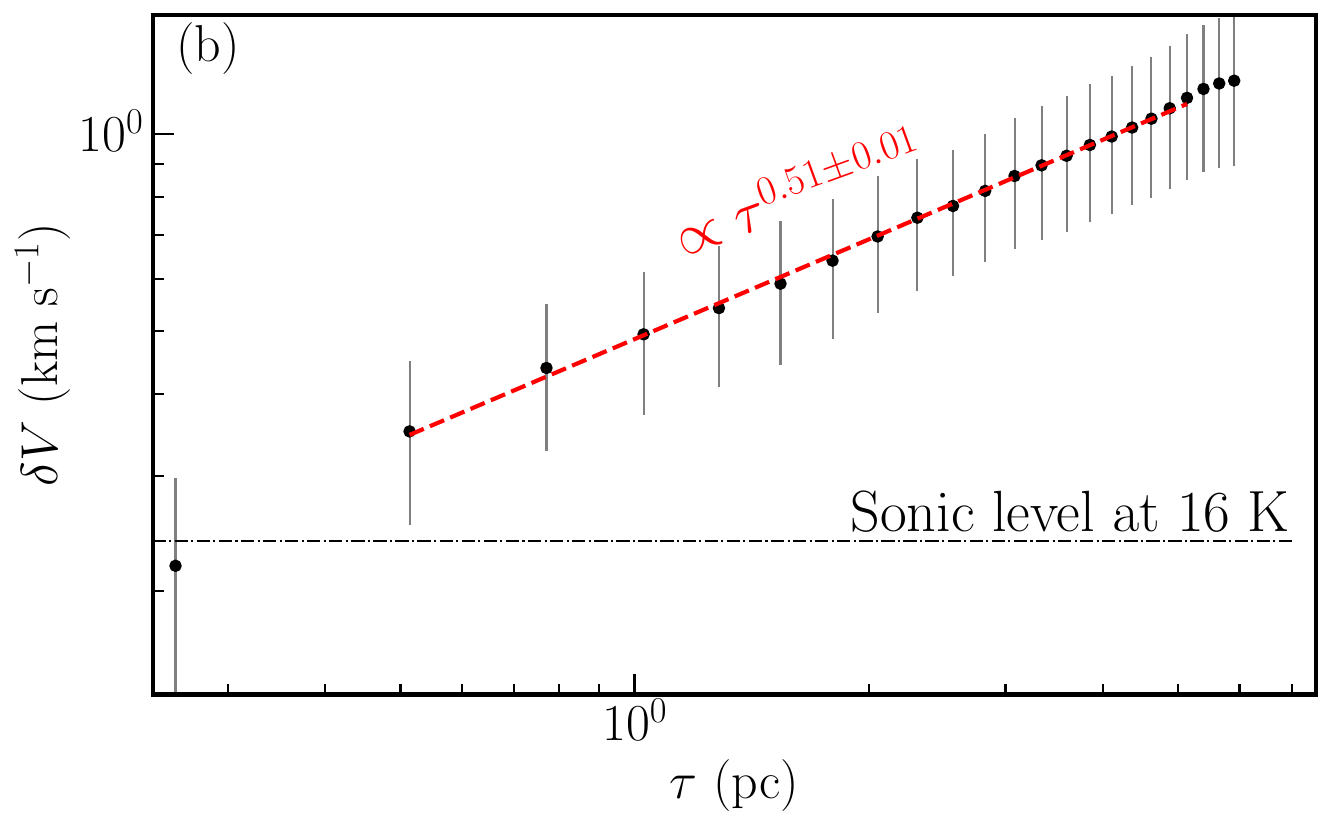}
\includegraphics[width=0.32\textwidth, height=0.23\textwidth]{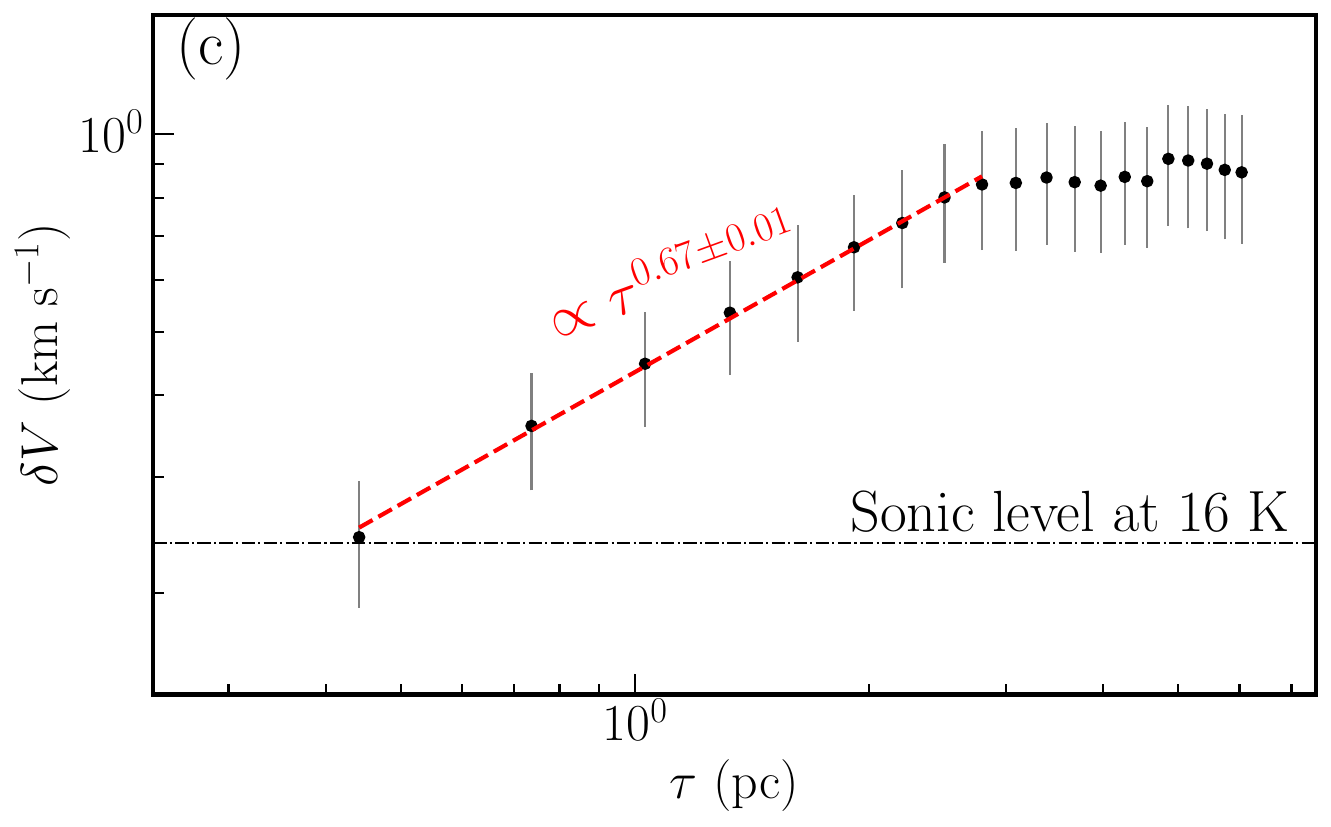}
\caption{(a): ACF of moment\,1 map of \etco~(2-1) of G34. The effective, positive correlation of the velocity deviation  ($\delta V$) with spatial scale ($\tau$) is constrained between 0.5\,pc and 5.4\,pc (see two dashed lines), where the former limit corresponds to an observing resolution of 0.5\,pc and the latter provides the largest effective scale of coherent velocity deviation (i.e., $C(\tau)=0$). (b):  Same as panel\,a but for  second-order SF analysis along the filamentary direction. (c): same as panel\,b but for analysis over all direction of the G34 cloud. The red lines indicate the best fit of a power-law within the effective range of correlation scales as derived in panel\,a. The horizontal line gives a sonic level of 0.24\,\vel.   In all panels, the error bar represents the standard error of the mean.
}
\label{fig:ACF_SF}
\end{figure*}

\begin{figure}
\centering
\includegraphics[width=0.48\textwidth]{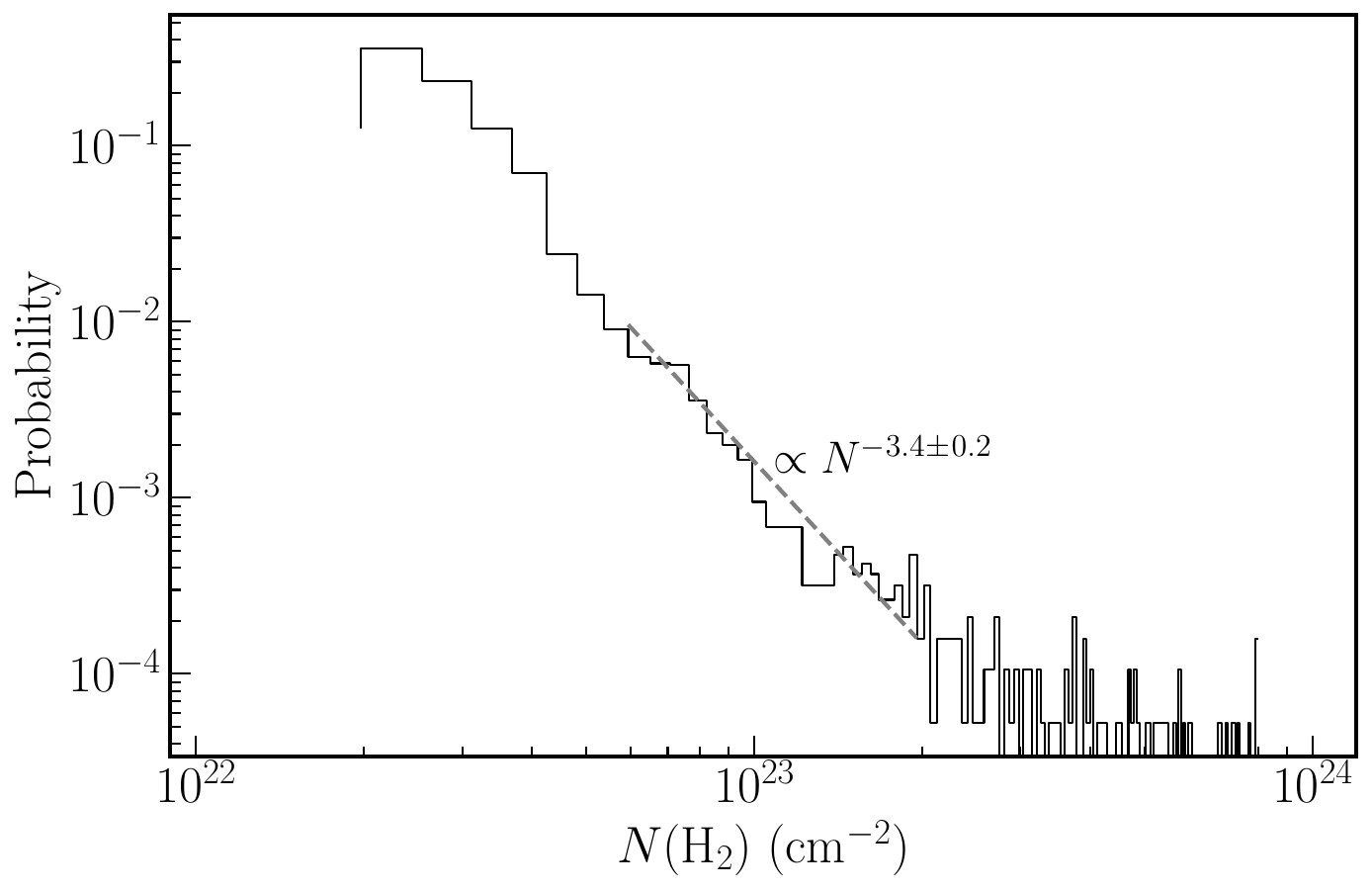}
\caption{  Column density distribution function (N-PDF) of the entire G34 cloud. The dashed line represents the best power-law fit in the high density regime between $2 N_{\rm bg} ({\rm H_2}$) and $3\times10^{23}$\,\pcmsq, beyond which sample points appear insufficient for the fit. $N_{\rm bg}$ is equal to $2.7\times10^{22}$\,\pcmsq.}
\label{fig:NPDF}
\end{figure}

Numerous observational and numerical studies are converging to a multiscale dynamical scenario of HMSF where filaments are of critical importance \citep{Beu18,Vaz19,Pad20,Kum20,Liu22b,Liu23,Sah22}.  In this scenario, filaments act as a `conveyor belt' to transport gas material between larger and smaller scales \egcite{Vaz19,Pad20}.

In \filname, the filament-aligned, coherent velocity gradients in the southern $F_{\rm S}$ and northern $F_{\rm N}$ filaments indicates the presence of large-scale flows onto the middle ridge of dense clumps. In the $F_{\rm S}$ filament, the longitudinal flows could be directed onto the middle ridge of the G34 cloud due to no appearance of other dense clumps condensed out of $F_{\rm S}$ (see Fig.\,\ref{fig:pv_diagram}). In contrast, only portions of the longitudinal flows in the $F_{\rm N}$ filament could feed the middle ridge due to the competition for gas partition from other dense clumps (i.e., MM3, and MM\,6--9) embedded in $F_{\rm N}$. Except for MM3, these clumps have masses around an order of magnitude lower than those in the middle ridge, likely rendering negligible influence on the gas flows onto the middle ridge. Regardless of MM3,  the $F_{\rm N}$ filament only between MM1 and MM9 could provide gas flows onto the middle ridge. 

Moving down to massive clumps in the ridge, MM5 and MM1 could be receiving gas feeding from $F_{\rm S}$-- and $F_{\rm N}$--rooted flows, respectively, since they are located at the southern and northern ends of the ridge. This possibility is  particularly supported in MM5 where the clump-scale infall rate is comparable to that of the southern filament-aligned flows at an order of $10^{-4}$\,\mdotyr, indicating the coherent gas flows from filament to clump. Additionally, both MM2 and MM4 clumps in the ridge could be fed by gas flows from both $F_{\rm S}$ and $F_{\rm N}$ filaments. This is because the gas infall rate of both clumps can become comparable to the filament-scale inflow rate at an order of $10^{-3}$\,\mdotyr\ until  the gas inflows of two filaments are combined along with uncertainties considered in Sect.\,3.1. 
According to \citet{Liu22a}, the clump-scale gas infall rates of MM2, MM4, and MM5 could be at least an order of magnitude higher than on smaller-scale cores. Particularly, the core-scale infall rate within MM1 is approximately at an order of $10^{-5}$\,\mdotyr--$10^{-4}$\,\mdotyr\ estimated from ALMA observations \citep{Liu22a}. Note that this estimate could be a lower limit since it was indirectly derived from the outflow rate of embedded young stars without accounting for the ionized gas contribution and the exact outflow inclination angle.

From the multiscale signatures of gas inflow/infall on different structures from filamentary clouds, clumps to cores, we confirm the multiscale dynamical scenario of HMSF being at work in G34 as proposed by \citet{Liu22a}, where filaments function as a `conveyor belt' to transport  their mass through clumps and cores down to protostars in a top-down manner.

\subsection{Driving mechanism of the multiscale HMSF scenario: gravity or turbulence?}

 The driving mechanism behind the multiscale HMSF scenario remains a topic of debate in the latest generation of HMSF models, such as GHC and I2. For instance, GHC posits that gravitational contraction of clouds is the driver across all scales, from large-scale clouds to the smallest seeds of star formation \citep{Vaz19}. On the other hand, I2 suggests that supersonic turbulent flows, rather than cloud self-gravity, are responsible on the largest cloud scale \citep{Pad20}. This discrepancy necessitates further dedicated investigations.

In G34, on small clump scales,  the gas infall associated with the multiscale HMSF scenario should be driven by gravity, which agrees with the subvirial state ($\alpha_{\rm vir}<2$) of the middle ridge of clumps (see Sect.\,3.2). Likewise, the gravity drives gas infall on smaller-scale cores due to their subvirial state (see table\,1 of \citealt{Liu22a}).

 On the filament scale, the influence of gravity can be inferred from the column density probability density function (N-PDF, \citealt{Vaz01, Bal11}). Theoretically, a purely lognormal distribution is expected if the density structure is shaped by supersonic, isothermal turbulence. However, deviations in the form of a power law for high column densities are predicted for self-gravitating clouds \citep{Bal11}. Fig.\,\ref{fig:NPDF} shows the N-PDF of the entire G34 cloud, which hosts the main G34 filament. In the high-density regime between $2 N_{\rm bg} ({\rm H_2}$) and $3\times10^{23}$\,\pcmsq\, beyond which sample points seem insufficient for analysis, the N-PDF best fits a slope of $m\sim-3.4$. This slope corresponds to an equivalent spherical density profile of power $\alpha$ in the form $\rho \propto r^{-\alpha}$ with $\alpha=-2/m+1$ \citep{Fed12, Sch13}. Assuming spherical symmetry is a rough approximation for complex and large regions. Yet, if the pixels contributing to the power-law form belong to a single condensation, this approximation is relatively appropriate. A spherical self-gravitating cloud has an $\alpha$ value between 1.5 and 2 \egcite{Sch13, Sch15}. Thus, the $\alpha \sim 1.6$ value suggests gravity's significance in regions with column densities above $2 N_{\rm bg} ({\rm H_2})$, which encompasses the main G34 filament structure as shown in contours in Figs.\,\ref{fig:filoverview} and \ref{fig:m1_c18O}. This implies that the associated filament-scale gas inflows responsible for HMSF in G34 could be gravity-driven.  Hereto, the gas dynamics or density distribution observed from filaments down to cores support evidence of self-gravitating density structures on these multiple scales, which agrees with the general finding by \citealt{Per23} from a systematic dynamics study toward  27 IRDC clouds.

 Interactions between turbulence and gravity happen everywhere in molecular clouds and thus regulate gas dynamics therein.
Turbulent motions of molecular clouds are known to follow the empirical Larson's relation, $\delta V \propto \tau^{\gamma}$, where $\delta V$ is the velocity deviation over the cloud size $\tau$, and the index $\gamma$ is widely adopted to be $\sim 0.5$ \egcite{Lar81, Sol87,Hey04}.  
The velocity deviation can  be quantified either by the line width or velocity dispersion \egcite{Lar81, Sol87}, or by the variance of the velocity centroid \egcite{Hey04}. The latter method has the advantage of not being affected by the broadening of the observed line width due to line-of-sight signal integration \egcite{Liuj23}.
Therefore, the Larson's velocity-size relation is investigated here using the  second-order structure function (SF) and autocorrelation function (ACF) of velocity centroids of molecular clouds.
Here, ACF analysis provides the largest effective scale of turbulence, $\tau_{\rm max}$, within which the velocity deviation is correlated positively with the spatial scale. Accordingly, robust SF analysis can be conducted within a spatial range of [$\tau_{\rm min}$,  $\tau_{\rm max}$], where $\tau_{\rm min}$ is determined by the observing resolution of 0.5\,pc. 
Note that $\tau_{\rm max}$ should be in principle less than half of the image size (i.e., $<5.9$\,pc in our case); otherwise some parts of the image will be shifted out of the frame and not be compared with the original image in ACF and SF analysis.

In Fig.\,\ref{fig:ACF_SF}a, we present the ACF plot of the G34 cloud calculated from the moment\,1 map of the \etco~(2--1) de-noised cube along the Dec. axis, which represents the south-north filamentary elongation of the whole cloud very well (see Fig.\,\ref{fig:filoverview}). The ACF plot gives $\tau_{\rm max}=5.4$\,pc (i.e., $C(\tau)=0$), below which $\delta V$ shows a strong positive correlation with $\tau$. Fig.\,\ref{fig:ACF_SF}b shows the same plot but for SF analysis. A power-law function is fitted within the spatial range of [$\tau_{\rm min}$, $\tau_{\rm max}$] to $\delta V$, resulting in a best-fit slope of $\sim 0.5$. This slope could be caused either by the large-scale velocity gradients, or by turbulent motions in filaments. The former possibility is expected to be higher on large scales (e.g., $>1$\,pc larger than clump scales in our case) since the filament-aligned velocity gradients reflect the large-scale dynamics of the G34 cloud. However, this possibility should be considered in practice rather low since the same SF analysis for such velocity gradients on scales of $>1$\,pc gives a slope of $\sim 0.2$ (see Fig.\,\ref{fig:pv_diagram}b), which deviates significantly from 0.5 as requested by Larson's relation.  Therefore, the slope of $\sim 0.5$ mostly arises from supersonic turbulent motions. Their supersonic nature can be inferred from their velocity deviations above a sonic level of 0.24\,\vel\ (see  Fig.\,\ref{fig:ACF_SF}b), at an average temperature of $\sim 16$\,K of the G34 cloud, which was estimated from its dust temperature map (see Sect.\,2.1).  It is worth noting that SF analysis in Fig.\,\ref{fig:pv_diagram}b refers to the dense filamentay structure with a width of $0.6$\,pc while Fig.\,\ref{fig:ACF_SF}b points to same analysis but for the entire cloud with diffuse cloud gas included. This fact results in the different SF distributions along the filamentary direction in both figures.

If the turbulence responsible for dynamics of the G34 filamentary cloud is isotropic, the SF analysis over all directions would result in the slope close to that (i.e., $\gamma\sim 0.5$) derived from the same analysis but only along the filamentary direction. Fig.\,\ref{fig:ACF_SF}c presents the average SF analysis over all directions. It yields a much steeper slope of $\gamma\sim 0.7$ than that inferred from the specific direction only along the filamentary structure. As reported in previous studies \egcite{Hey12,Ott17,Luk22}, this difference would be attributed to the anisotropic nature of turbulence in interstellar medium, for example, along the two orthogonal directions of the G34 filament.

Turbulent motions within the entire G34 cloud, which includes the major dense filament, could originate from several sources such as gravity and cloud-cloud collisions. As previously discussed, gravity is the dominant force on the filament scale within the cloud. This could drive the gas inflows aligned with the filament onto the central ridge of massive clumps. This mechanism, driven by gravity, involves the conversion of gravitational energy into kinetic energy stored in turbulent motions. Moreover, the presence of three subfilaments in the G34 cloud and an associated east-west velocity gradient across the northern cloud suggest cloud-cloud collisions among the subfilaments. These collisions could be another source of turbulent motions on the entire cloud scale. In addition to gravity and cloud-cloud collisions, large-scale injections such as compression via supernova explosions or expanding \hii\ regions could contribute to turbulent motions on the G34 cloud scale. In this context, the \hii\ region adjacent to the southern G34 cloud, referred to as G34.172+0.175 \citep{And11} and seen as an IR bubble in 8\,\um\ emission, could be a contributing factor.

 Overall, we suggest that in the G34 cloud, gravity governs gas dynamics and the associated HMSF process on multiscales up to the filament scale; on the largest cloud scale, turbulence becomes important, which could be driven by several sources, including the self-gravity of the G34 cloud itself.

\section{Conclusion}
We have investigated multiscale gas dynamics related to high-mass star formation (HMSF) in the filamentary G34 cloud, using APEX observations of molecular lines of \etco~(2-1), \hco/\hthco~(3-2), and HCN/\hthcn~(3-2). The major findings are as follows.

Large-scale, filament-aligned velocity gradients from \etco\ emission are detected in the southern $F_{\rm S}$ and northern $F_{\rm N}$ filamentswhich drive filamentary gas inflows onto dense clumps in the middle ridge of G34.
The total mass inflow rate of both filaments is approximately $\sim 5.5\times10^{-4}$\,\mdotyr, which would be doubled if the remaining gas inflowing off the filaments were considered. The origin of these filament-aligned inflows could be driven by gravity, as reflected from a steep power-law form of the N-PDF distribution in the high density regime above $2.7\times10^{22}$\,\pcmsq.  Nevertheless, on the scale of the entire cloud that encompasses the main G34 filament, turbulent motions are revealed. These motions, driven by various sources including the cloud’s self-gravity, are manifested in the structure function of the velocity centroids of the \etco\ line.

Clump-scale gas infalls are observed in MM2, MM4, and MM5 clumps in the middle ridge, as characterized by asymmetric \hco/\hthco\ line profiles with brighter emission on the blue side of the clump’s systemic velocity. The gas infall rates are  estimated to be $5.7\pm1.4\times10^{-3}$\,\mdotyr\ for MM2, $2.7\pm0.6\times10^{-3}$\,\mdotyr\ for MM4, and $4.4\pm1.3\times10^{-4}$\,\mdotyr\ for MM5. They could depend on filament-aligned gas inflows since the infall/inflow rates on these two scales are comparable. Gravity is found to be the major driver of gas infall on smaller scales of clumps and cores, which agrees with their subvirial state ($\alpha_{\rm vir}$<2).

 In conclusion, the signatures of the multiscale gas inflows/infalls on different scales, from large-scale filamentary clouds, clumps to cores where stars form, provide strong observational evidence for the multiscale dynamical HMSF scenario in G34.  This multiscale scenario could be driven by gravity up to the filament scale, Beyond that, turbulence could be at work in G34, driven by several sources including gravity.

%%%%%%%%%%%%%%%%%%%%%%%%%%%%%%%%%%%%%%%%%%%
\section*{Acknowledgements}
 We thank the anonymous referee for comments and suggestions that greatly improved the quality of this paper. 
This work has been supported by the National Key R\&D Program of China (No.\,2022YFA1603101). H.-L. Liu is supported by National Natural Science Foundation of China (NSFC) through the grant No.\,12103045, by Yunnan Fundamental Research Project (grant No.\,202301AT070118), and by Xingdian Talent Support Plan--Youth Project. S.-L. Qin is supported by NSFC under No.\,12033005.

\bibliographystyle{aasjournal}

\begin{thebibliography}{}
\expandafter\ifx\csname natexlab\endcsname\relax\def\natexlab#1{#1}\fi
\providecommand{\url}[1]{\href{#1}{#1}}

\bibitem[Anderson et al.(2011)]{And11} Anderson, L.~D., Bania, T.~M., Balser, D.~S., et al.\ 2011, \apjs, 194, 32. doi:10.1088/0067-0049/194/2/32
\bibitem[Avison et al.(2021)]{Avi21} Avison, A., Fuller, G.~A., Peretto, N., et al.\ 2021, \aap, 645, A142. doi:10.1051/0004-6361/201936043
\bibitem[Ballesteros-Paredes et al.(2011)]{Bal11} Ballesteros-Paredes, J., V{\'a}zquez-Semadeni, E., Gazol, A., et al.\ 2011, \mnras, 416, 1436. doi:10.1111/j.1365-2966.2011.19141.x
\bibitem[Beuther et al.(2018)]{Beu18} Beuther, H., Mottram, J.~C., Ahmadi, A., et al.\ 2018, \aap, 617, A100. doi:10.1051/0004-6361/201833021
\bibitem[Bonnell et al.(2001)]{Bon01} Bonnell, I.~A., Bate, M.~R., Clarke, C.~J., et al.\ 2001, \mnras, 323, 785. doi:10.1046/j.1365-8711.2001.04270.x
\bibitem[Clarke et al.(2018)]{Cla18} Clarke, S.~D., Whitworth, A.~P., Spowage, R.~L., et al.\ 2018, \mnras, 479, 1722. doi:10.1093/mnras/sty1675
\bibitem[De Vries \& Myers(2005)]{De 05} De Vries, C.~H. \& Myers, P.~C.\ 2005, \apj, 620, 800. doi:10.1086/427141
\bibitem[Federrath \& Klessen(2012)]{Fed12} Federrath, C. \& Klessen, R.~S.\ 2012, \apj, 761, 156. doi:10.1088/0004-637X/761/2/156
\bibitem[Guilloteau \& Lucas(2000)]{Gui00} Guilloteau, S. \& Lucas, R.\ 2000, Imaging at Radio through Submillimeter Wavelengths, 217, 299
\bibitem[G{\"u}sten et al.(2006)]{Gus06} G{\"u}sten, R., Nyman, L. {\r{A}}., Schilke, P., et al.\ 2006, \aap, 454, L13. doi:10.1051/0004-6361:20065420
\bibitem[Heyer \& Brunt(2004)]{Hey04} Heyer, M.~H. \& Brunt, C.~M.\ 2004, \apjl, 615, L45. doi:10.1086/425978
\bibitem[Heyer \& Brunt(2012)]{Hey12} Heyer, M.~H. \& Brunt, C.~M.\ 2012, \mnras, 420, 1562. doi:10.1111/j.1365-2966.2011.20142.x
\bibitem[He et al.(2023)]{HeY23} He, Y.-X., Liu, H.-L., Tang, X.-D., et al.\ 2023, \apj, 957, 61. doi:10.3847/1538-4357/acf766
\bibitem[Kauffmann et al.(2013)]{Kau13} Kauffmann, J., Pillai, T., \& Goldsmith, P.~F.\ 2013, \apj, 779, 185. doi:10.1088/0004-637X/779/2/185
\bibitem[Kirk et al.(2013)]{Kir13} Kirk, H., Myers, P.~C., Bourke, T.~L., et al.\ 2013, \apj, 766, 115. doi:10.1088/0004-637X/766/2/115
\bibitem[Krumholz et al.(2005)]{Kru05} Krumholz, M.~R., McKee, C.~F., \& Klein, R.~I.\ 2005, \nat, 438, 332. doi:10.1038/nature04280
\bibitem[Kumar et al.(2020)]{Kum20} Kumar, M.~S.~N., Palmeirim, P., Arzoumanian, D., et al.\ 2020, \aap, 642, A87. doi:10.1051/0004-6361/202038232
\bibitem[Larson(1981)]{Lar81} Larson, R.~B.\ 1981, \mnras, 194, 809. doi:10.1093/mnras/194.4.809
\bibitem[Liu et al.(2018)]{Liu18} Liu, H.-L., Stutz, A., \& Yuan, J.-H.\ 2018, \mnras, 478, 2119. doi:10.1093/mnras/sty1270
\bibitem[Liu et al.(2019)]{Liu19} Liu, H.-L., Stutz, A., \& Yuan, J.-H.\ 2019, \mnras, 487, 1259. doi:10.1093/mnras/stz1340
\bibitem[Liu et al.(2020)]{Liu20} Liu, H.-L., Sanhueza, P., Liu, T., et al.\ 2020, \apj, 901, 31. doi:10.3847/1538-4357/abadfe
\bibitem[Liu et al.(2021)]{Liu21} Liu, H.-L., Liu, T., Evans, N.~J., et al.\ 2021, \mnras, 505, 2801. doi:10.1093/mnras/stab1352
\bibitem[Liu et al.(2022a)]{Liu22a} Liu, H.-L., Tej, A., Liu, T., et al.\ 2022a, \mnras, 510, 5009. doi:10.1093/mnras/stab2757
\bibitem[Liu et al.(2022b)]{Liu22b} Liu, H.-L., Tej, A., Liu, T., et al.\ 2022b, \mnras, 511, 4480. doi:10.1093/mnras/stac378
\bibitem[Liu et al.(2023)]{Liu23} Liu, H.-L., Tej, A., Liu, T., et al.\ 2023, \mnras, 522, 3719. doi:10.1093/mnras/stad047
\bibitem[\protect\citeauthoryear{Liu et al.}{2023}]{Liuj23} Liu J., Zhang Q., Liu H.~B., Qiu K., Li S., Li Z.-Y., Ho P.~T.~P., et al., 2023, ApJ, 949, 30. doi:10.3847/1538-4357/acc4c0
\bibitem[Mai et al.(2023)]{Mai23} Mai, X., Zhang, B., Reid, M.~J., et al.\ 2023, \apj, 949, 10. doi:10.3847/1538-4357/acc52a
\bibitem[Luk et al.(2022)]{Luk22} Luk, S.-S., Li, H.-. bai ., \& Li, D.\ 2022, \apj, 928, 132. doi:10.3847/1538-4357/ac574c
\bibitem[McKee \& Tan(2003)]{McK03} McKee, C.~F. \& Tan, J.~C.\ 2003, \apj, 585, 850. doi:10.1086/346149
\bibitem[Padoan et al.(2020)]{Pad20} Padoan, P., Pan, L., Juvela, M., et al.\ 2020, \apj, 900, 82. doi:10.3847/1538-4357/abaa47
\bibitem[Otto et al.(2017)]{Ott17} Otto, F., Ji, W., \& Li, H.-. bai .\ 2017, \apj, 836, 95. doi:10.3847/1538-4357/836/1/95
\bibitem[Pan et al.(2023)]{Pan23} Pan, S., Liu, H.-L., \& Qin, S.-L.\ 2023, \mnras, 519, 3851. doi:10.1093/mnras/stac3658
\bibitem[Peretto et al.(2013)]{Per13} Peretto, N., Fuller, G.~A., Duarte-Cabral, A., et al.\ 2013, \aap, 555, A112. doi:10.1051/0004-6361/201321318
\bibitem[Peretto et al.(2023)]{Per23} Peretto, N., Rigby, A.~J., Louvet, F., et al.\ 2023, arXiv:2305.02701. doi:10.48550/arXiv.2305.02701
\bibitem[Rathborne et al.(2005)]{Rat05} Rathborne, J.~M., Jackson, J.~M., Chambers, E.~T., et al.\ 2005, \apjl, 630, L181. doi:10.1086/491656
\bibitem[Rathborne et al.(2006)]{Rat06} Rathborne, J.~M., Jackson, J.~M., \& Simon, R.\ 2006, \apj, 641, 389. doi:10.1086/500423
\bibitem[Saha et al.(2022)]{Sah22} Saha, A., Tej, A., Liu, H.-L., et al.\ 2022, \mnras, 516, 1983. doi:10.1093/mnras/stac2353
\bibitem[Sanhueza et al.(2010)]{San10} Sanhueza, P., Garay, G., Bronfman, L., et al.\ 2010, \apj, 715, 18. doi:10.1088/0004-637X/715/1/18
\bibitem[Schneider et al.(2013)]{Sch13} Schneider, N., Andr{\'e}, P., K{\"o}nyves, V., et al.\ 2013, \apjl, 766, L17. doi:10.1088/2041-8205/766/2/L17
\bibitem[Schneider et al.(2015)]{Sch15} Schneider, N., Csengeri, T., Klessen, R.~S., et al.\ 2015, \aap, 578, A29. doi:10.1051/0004-6361/201424375
\bibitem[Solomon et al.(1987)]{Sol87} Solomon, P.~M., Rivolo, A.~R., Barrett, J., et al.\ 1987, \apj, 319, 730. doi:10.1086/165493
\bibitem[Tang et al.(2019)]{Tan19} Tang, Y.-W., Koch, P.~M., Peretto, N., et al.\ 2019, \apj, 878, 10. doi:10.3847/1538-4357/ab1484
\bibitem[V{\'a}zquez-Semadeni \& Garc{\'\i}a(2001)]{Vaz01} V{\'a}zquez-Semadeni, E. \& Garc{\'\i}a, N.\ 2001, \apj, 557, 727. doi:10.1086/321688
\bibitem[V{\'a}zquez-Semadeni et al.(2019)]{Vaz19} V{\'a}zquez-Semadeni, E., Palau, A., Ballesteros-Paredes, J., et al.\ 2019, \mnras, 490, 3061. doi:10.1093/mnras/stz2736
\bibitem[V{\'a}zquez-Semadeni et al.(2023)]{Vaz23} V{\'a}zquez-Semadeni, E., G{\'o}mez, G.~C., \& Gonz{\'a}lez-Samaniego, A.\ 2023, arXiv:2306.13846. doi:10.48550/arXiv.2306.13846
\bibitem[Xu et al.(2023)]{Xu 23} Xu, F.-W., Wang, K., Liu, T., et al.\ 2023, \mnras, 520, 3259. doi:10.1093/mnras/stad012
\bibitem[Yang et al.(2023)]{Yan23} Yang, D., Liu, H.-L., Tej, A., et al.\ 2023, arXiv:2306.10332. doi:10.48550/arXiv.2306.10332
\bibitem[Yuan et al.(2018)]{Yua18} Yuan, J., Li, J.-Z., Wu, Y., et al.\ 2018, \apj, 852, 12. doi:10.3847/1538-4357/aa9d40
\bibitem[Zucker \& Chen(2018)]{Zuc18} Zucker, C. \& Chen, H.~H.-H.\ 2018, \apj, 864, 152. doi:10.3847/1538-4357/aad3b5
\end{thebibliography}

\appendix
\section{Radial density profile of the G34 filament} 
\label{app:radial_profile}

\begin{figure}
\centering
\includegraphics[width=0.48\textwidth]{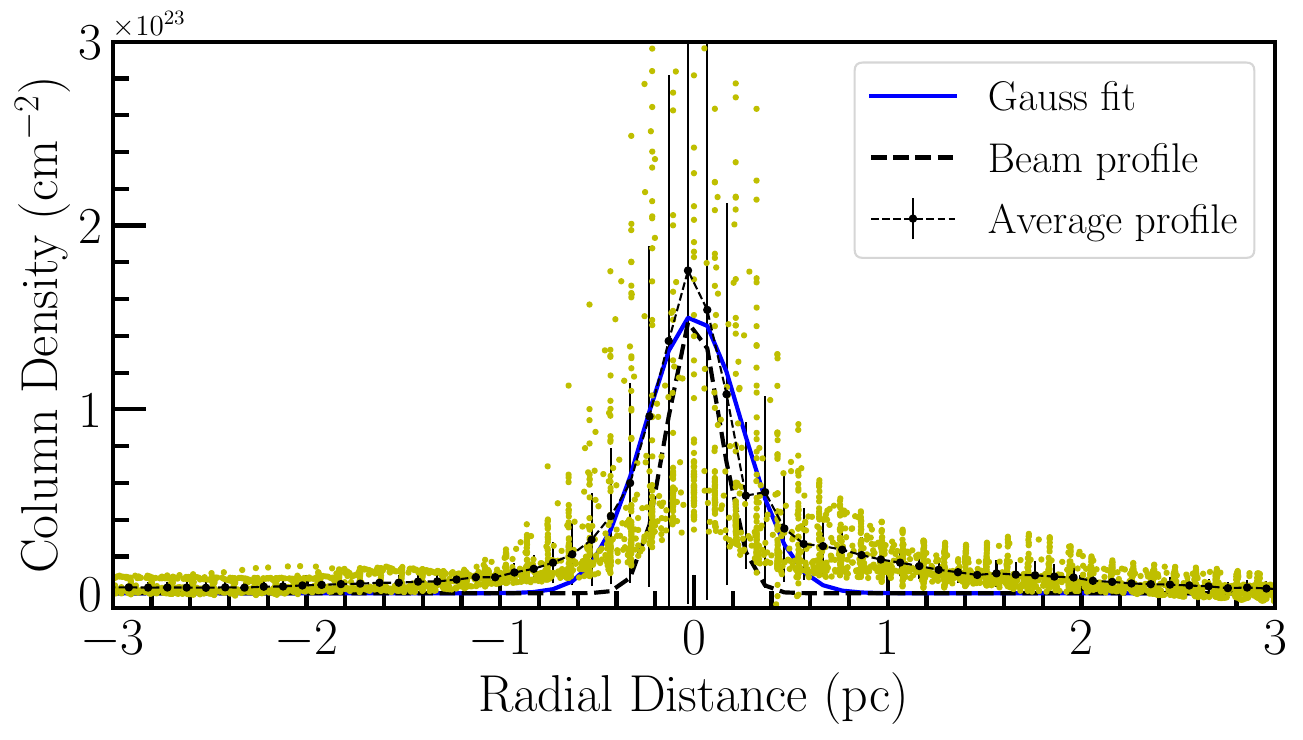}
\caption{ Radial column density profiles (yellow dots) of the G34 filament, along with
the mean density profile (blue dots with error bars). All of the density profiles are the results from a background subtraction of $N_{\rm bg}({\rm H_{2}}) = 2.7 \times 10^{22}$\,\pcmsq. 
The blue solid line represents the best Gaussian fit while 
the gray dashed line corresponds to the $18\arcsec$-beam Gaussian profile.}
\label{fig:NH2_radial_profile}
\end{figure}

 We used the Python package, RadFil \citep{Zuc18}, to identify the filamentary structure. This tool extracts the ridgeline from the $18\arcsec$-resolution column density ($N{\rm (H_2)}$) map by pinpointing the peak value along the filament (see Fig.\,\ref{fig:filoverview}). We also created density profiles along the filament ridge (see Fig.\,\ref{fig:NH2_radial_profile}) as below. For positions sampled every three pixels (equivalent to one beam size of the column density map) along the filament ridge, we generated individual column density profiles in the local tangent direction. The mean column density profile was then computed by averaging these individual profiles across all positions. 
We fitted the mean density profile with a Gaussian function (see Fig.\,\ref{fig:NH2_radial_profile}). The fit resulted in a deconvolved FWHM of $\sim 0.6$\,pc and a background column density of $\sim 2.7\times10^{22}$\,\pcmsq.

\begin{figure}
\centering
\includegraphics[width=0.8\textwidth, height=0.8\textwidth]{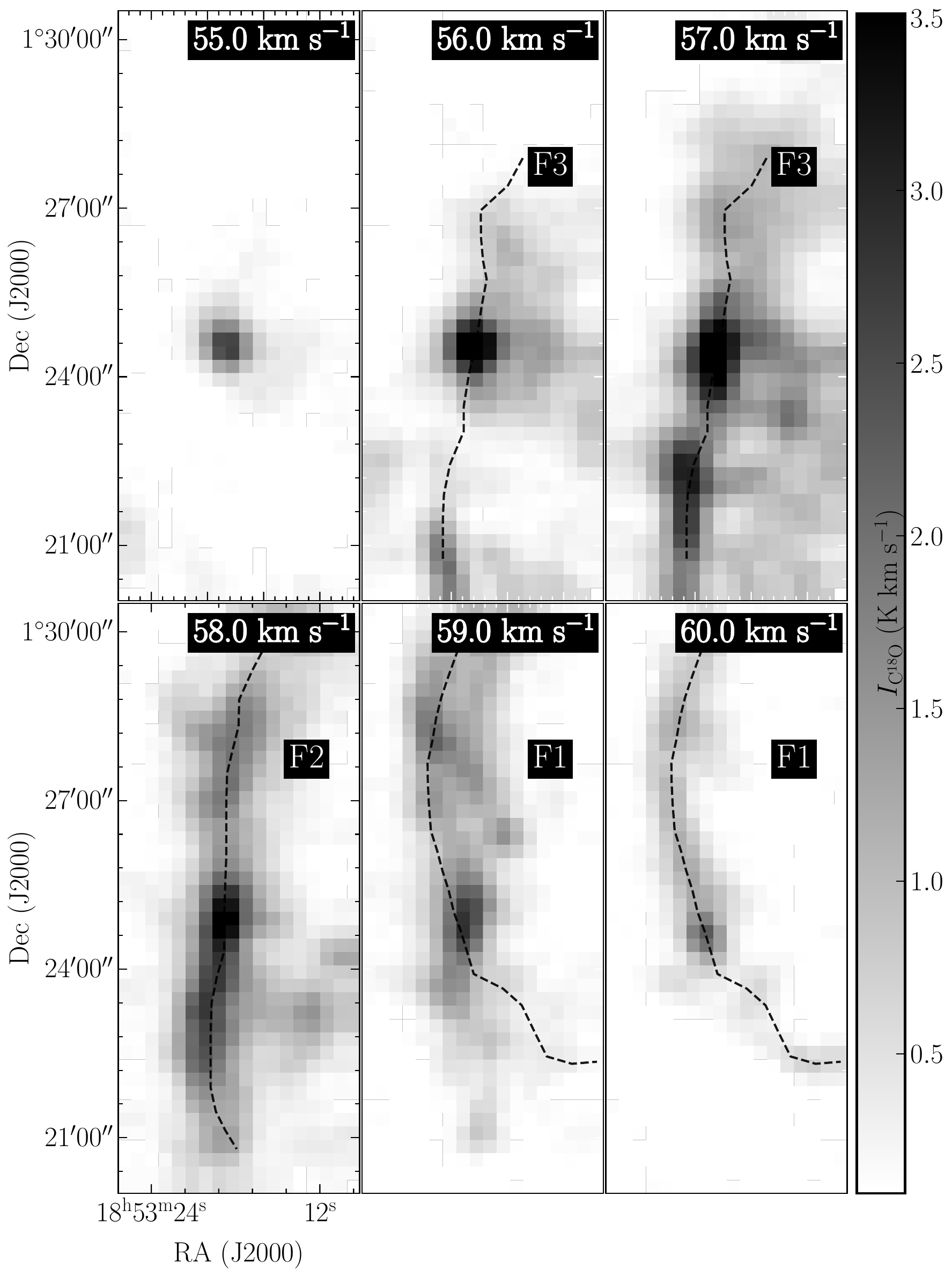}
\caption{ Velocity channel map of \etco~(2--1) of the G34 cloud. Dashed curves represent velocity-coherent subfilaments identified by eye from the channel map.}
\label{fig:vel_chan}
\end{figure}

\begin{figure}[!ht]
\centering
\includegraphics[width=0.45\textwidth]{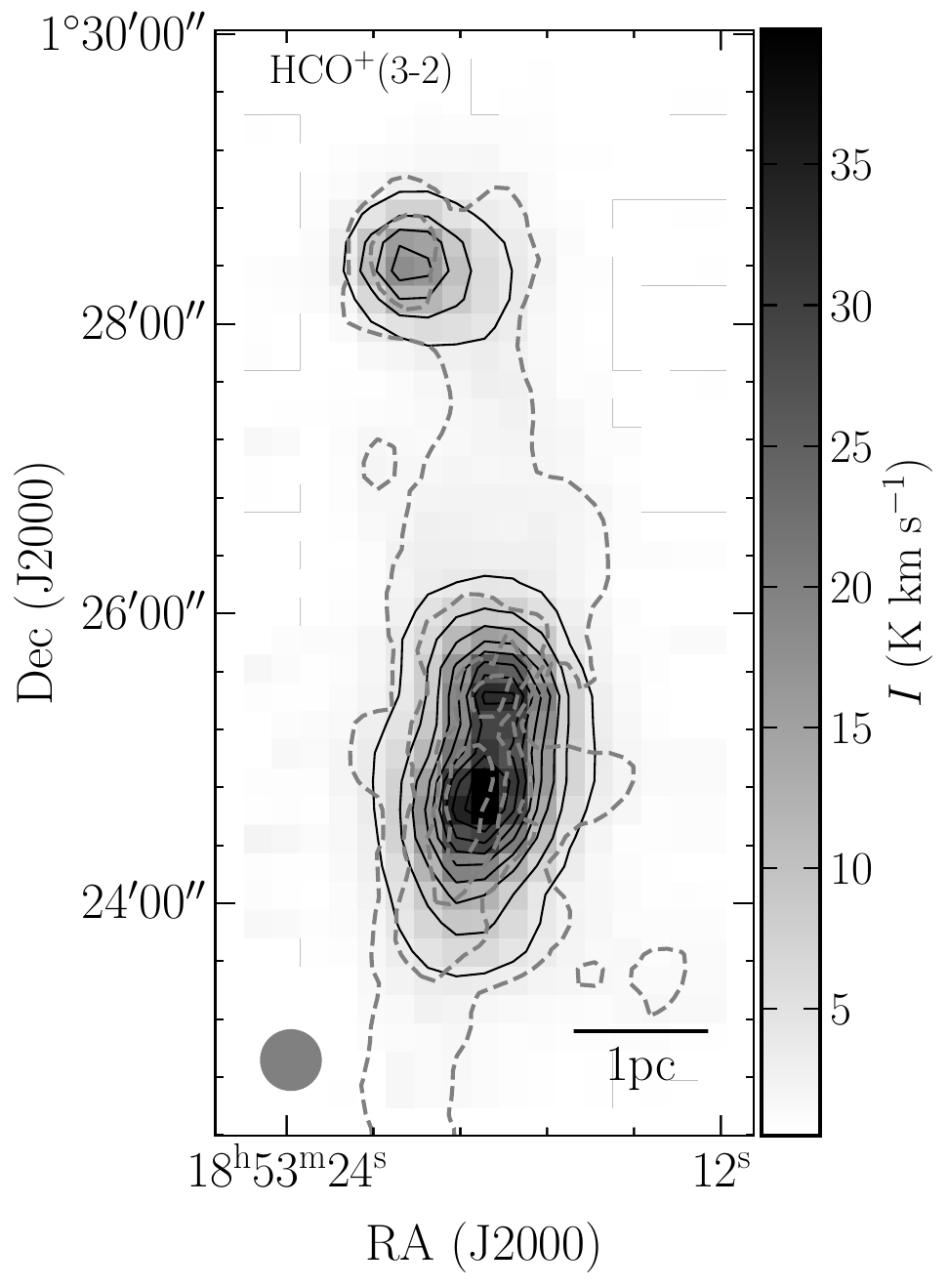}
\caption{  Velocity-integrated intensity map of \hco~(3--2) overlaid with the H$_2$ column density contours. The contour levels are the same as those in Fig.\,\ref{fig:filoverview}.
}
\label{fig:M0_h13cop}
\end{figure}

\end{document}